\documentclass[12pt,preprint]{aastex}
\usepackage{amsmath}
\usepackage{graphicx}
\usepackage{amsmath,bm}
\usepackage{multirow}
\usepackage{units}
\DeclareUnicodeCharacter{2212}{-}
\usepackage{color}

\begin{document}
\begin{tiny}

\end{tiny}

\title{Do the periodic activities of repeating fast radio bursts represent the spins of neutron stars?}

\author{Kun Xu$^{1,2}$, Qiao-Chu Li$^{3,4}$, Yuan-Pei Yang$^{5}$, Xiang-Dong Li$^{3,4}$, Zi-Gao Dai$^{6,3,4}$ and Jifeng Liu$^{1,2,7}$}

\affil{$^{1}$School of Astronomy and Space Sciences, University of Chinese Academy of Sciences, Beijing, People’s Republic of China; xukun@smail.nju.edu.cn}

\affil{$^{2}$Key Laboratory of Optical Astronomy, National Astronomical Observatories, Chinese Academy of Sciences, Beijing, People’s Republic of China}

\affil{$^{3}$Department of Astronomy, Nanjing University, Nanjing 210023, People’s Republic of China}

\affil{$^{4}$Key Laboratory of Modern Astronomy and Astrophysics, Nanjing University,
Ministry of Education, Nanjing 210023, People’s Republic of China}

\affil{$^{5}$South-Western Institute for Astronomy Research, Yunnan University, Kunming 650500, People’s Republic of China}

\affil{$^{6}$CAS Key Laboratory for Research in Galaxies and Cosmology, Department of Astronomy, University of Science and Technology of China, Hefei 230026, China}

\affil{$^{7}$WHU-NAOC Joint Center for Astronomy, Wuhan University, Wuhan, People’s Republic of China}

\begin{abstract}

Fast radio bursts (FRBs) are mysterious radio transients with millisecond durations. Recently, a periodic activity of $\sim$16 day and a possible periodicity of $\sim$159 day were detected to arise from FRB 180916.J0158+65 and FRB 121102, respectively, and the spin period of a slow-rotation magnetar was further considered to be one of possibilities to explain the periodic activities of repeating FRBs. For isolated neutron stars, the spin evolution suggests that it's difficult to reach several hours. In this work, we mainly
focus on the possible maximum spin period of isolated NSs / magnetars dominated by an interaction between star’s magnetic field and the disk.
We find that the disk wind plays an important role in spin evolution, whose influence varies the power law index in the evolution equation of mass flow rate.
For a magnetar without disk wind, the longest spin period is tens of hours. 
When the disk wind with a classical parameter is involved, the maximum spin period can reach hundreds of hours.
But for a much extremely large index of mass flow rate due to disk wind or other angular momentum extraction processes, a spin period of $\sim$(16-160) days is still possible.

\end{abstract}

\emph{Unified Astronomy Thesaurus concepts:} Accretion (14); Stellar accretion disks (1579); Magnetars (992); Radio bursts (1339);
Stellar rotation (1629)

\section{Introduction}

Fast radio bursts (FRBs) are radio transients with millisecond durations and extremely high brightness temperatures. Up to present, hundreds of FRBs have been discovered\footnote{See \citet{pet16} for a catalog of published FRBs (http://frbcat.org/) and the Transient Name Server system ( https://www.wis-tns.org/) for newly reported FRBs \citep{pet20}.} \citep[e.g.,][]{lor07,tho13,spi16,cha17,ban19,rav19,pro19,mar20}, and dozens of them are repeating sources. Although a Galactic FRB (FRB 200428) \citep{boc20,chi20b} was recently detected from a magnetar SGR J1935+2154 associated with a X-ray burst \citep{mer20,li20,tav20,rid20} during its active phase, the physical origin of FRBs is still poorly known \citep[e.g.,][]{kaz18, cor19, pet19, pla19}. Based on the distance of the emission region from the neutron star (NS), various models can be divided into ``close-in'' models and ``far-away'' models.
The ``close-in'' models proposed that FRBs are emitted from the magnetosphere of an NS due to internal triggers 
\citep[e.g.,][]{kum17,Lyutikov2017,kat18b,lu18,yan18,Wadiasingh2019,kum20,lu20,yan20a,wan20b,wan20,Lyutikov2020,wad20,Lyubarsky2020}
or external triggers \citep{zha17,dai20b,gen20}. And the ``far-away'' models suggested that FRBs are produced by synchrotron maser in a shocked outflow \citep{lyu14,bel20,wax17,met19,mar20,yu20,wu20,xia20}.

Remarkably, a periodic activity of $16.35\pm0.18\,\unit{day}$ was recently detected to arise from FRB 180916.J0158+65 (hereafter FRB 180916), and the activity window during each period is about 5 days \citep{Chime/Frb2020}. Very recently, with all detections of FRB 180916 from $110\,\unit{MHz}$ to $1765\,\unit{MHz}$ \citep{pas20, san20, Chime/Frb2020, mart20, agg20}, the period is confirmed to be $16.29_{-0.17}^{+0.15}\,\unit{day}$ and the activity window peaks earlier at higher frequencies \citep{pas20}. Furthermore, the first repeating FRB, FRB 121102, also appears tentative periodic activity of $\sim160$ days \citep{Rajwade2020, cru20}. Three possibilities were widely considered to explain the periodic activities of these sources \citep[see the review of][]{bz20}: 
1. the FRB source is in a binary system
containing an NS, and the observed period corresponds to the orbital period of the binary system \citep{iok20,lyu20,dai20,gu20,kue21,den21,Sridhar2021};
2. radio bursts are generated from a deformed NS with narrow beam, the observed periodic activity is due to the precession like a gyroscope  \citep{lev20,yan20,zan20,ton20,Li2021,Sridhar2021}; 
3. the observed period corresponds to an extremely slow rotation of an NS \citep{Beniamini2020}. 
The last possibility might be the simplest and most intriguing. 
If such a long period is due to the spin of an NS, one may have more opportunities to study the NS evolution \citep[e.g.,][]{Xu2019} or constrain the fundamental physics \citep[e.g.,][]{yan17}.
In this work, we will focus on the discussion of the last possibility, which equivalently raise the question: what is the longest possible spin period of an isolated NS?

In general, the long period of an NS could be due to magnetic dipole radiation or supernova fallback disk\footnote{In addition, gravitational wave emission \citep{Gittins2019} and neutrinos radiation 
can also lead to an NS spinning down, 
but their influences are too small to be considered, especially in slow spin systems.}. The former will extract the rotational energy via the magnetic dipole radiation and let the NS lose angular momentum.
In the latter case, the material in the fallback disk will be expelled if the disk is in propeller state, which will carry away the angular momentum of the NS.
In this work, we mainly
focus on the possible maximum spin period of isolated NSs / magnetars. 
Although some different mechanisms, like the winds or bursts/outbursts from magnetars \citep{Tong2020a,Beniamini2020}, may relax certain parameter requirements, a general and
detailed discussion about the most likely scenario, purely fallback, is still necessary,
especially for the discussion about the mysterious period of FRBs.

The pulsar 1E161348$-$5055 (hereafter 1E 1613), which has a spin period about 6.67 hr \citep{De2006}, is the slowest isolated NS known by now. 
To explain the extreme spin period of 1E 1613, previous studies show that an ultra-strong magnetic field of $\gtrsim 10^{15}$ G is not sufficient, an extra spin-down torque is also required, which is possibly delivered by a fallback disk \citep{De2006,Li2007,Tong2016,Ho2017}. 
\cite{Xu2019} (hereafter XL19) shows that 
the initial disk mass can only be in a very small range $\sim 10^{-7} M_\odot$.
Based on the model in XL19, we perform more general Monte-Carlo simulations to get the longest spin period of an isolated NS interacting with a fallback disk. 
Furthermore, we take disk wind into account and discuss its influence. 
We obtain the maximum spin period of an isolated NS is $\sim 63$ hr without considering disk wind and $\sim 13.5$ day with a classical parameter of disk wind.
While with a much extremely large parameter, which is the power law index of the mass flow rate due to disk wind or other angular momentum extraction process, we can obtain the spin period longer than 16-day or even 160-day.

This paper is organized as follows. We discuss the period constraints of an NS, in particular the fallback disk model in Section 2 and exhibit the results in Section 3. We discuss the range of disk wind parameter in Section 4 and summarize this paper in Section 5.

\section{Mechanisms Dominating Spin Evolution of Isolated NSs}

For an isolated NS without fallback disk, the magnetic dipole radiation gives a torque $N_{\rm B}=-2 \mu^2 \Omega^3/3 c^3$ on the NS.
which leads to the spin evolution as
\begin{equation}
    P_{dip}=\left(P_{\rm i}^2+\frac{16 \pi^2}{3 c^3 I} \mu^2 t \right)^{1/2} \approx 248 ~s \cdot B_{16} t_{4}^{1/2}.
\end{equation}
Where $P_{\rm i}$, $I$ and $\mu=BR_{\rm NS}^3$ are the initial spin period, moment of inertia and magnetic dipole moment of the NS, respectively. And $t_4$ is the age of the NS in units of $10^4$ yr.
We assume that the magnetic dipole radiation is exerted on the NS in its lifetime. 
Consequently, we can get the maximum spin period $P_{dip,\, \rm max} \lesssim 248$ s for a magnetar with $B \lesssim 10^{16}~{\rm G}$ and $t\lesssim10~{\rm kyr}$ if FRBs are from active magnetars.

As to the period of a magnetar which can generate FRBs, \cite{wad20} recently predicted a death line for magnetars in the framework of a low-twist magnetar magnetosphere with dislocations of magnetic foot-points inducing accelerating gaps.
When the density of the charge produced by the star motion to screen the electric field is insufficient, the intense particle acceleration would be produced.
And the death line satisfies $P_{low-twist} \gtrsim 6 ~\unit{ms} \cdot B^{-1}_{16}$,
which gives a lower limit of period that can generate FRBs from a magnetar. 
If $P_{\rm{dip}} > P_{\rm{low-twist}} $, it is possible for the magnetar to produce FRBs.

In conclusion, the longest period of an isolated NS/magnetar with the age $ < 10 \,\unit{kyr}$ cannot exceed several hundred seconds.
However, since the material in the fallback disk could be expelled by the rotating magnetosphere if the disk is in propeller state, the NS could further spin down, leading to a longer period. Meanwhile, the mass transfer in accretion process might trigger FRBs, which are generated by coherent curvature radiation of the electrons because of accretion material moving along the NS magnetic field lines \citep[e.g.,][]{gen20, gu20}.  
In the following part, we will focus on the period constraint from the fallback disk model.


In order to better understand the periodicity of isolated NSs, especially the long period, the fallback disk model is considered. 
A fallback disk may form around an NS if the angular momentum of the fallback material, which is gravitationally captured after the supernova explosion,
is sufficient \citep[e.g.,][]{wang2006}.
Then the disk evolves following the self-similar solutions of the viscous diffusion equation \citep{Pringle1981,Cannizzo1990}. 
We use the same model in \cite{Xu2019}, 
where the mass flow rate $\dot{m}$ (in units of Eddington accretion rate $\dot{M}_{\rm Edd}$) in the disk varies as \citep{Liu2015}
\begin{equation}
	\dot{m}=\dot{m_0} (t/t_{\rm f})^{-a}. \label{mdot}
\end{equation}
Here $\dot{m}_0=\dot{M}_0/\dot{M}_{\rm Edd}$ is the initial mass flow rate and $t_{\rm f}$ is the formation time of the fallback disk.
The power law index $a$ depends on opacity and disk wind loss in the slim disk phase, 
\begin{enumerate}
	\item $a=19/16$ for opacity dominated by electron scattering and $a=1.25$ for Kramers opacity \citep{Cannizzo1990,Francischelli2002,Li2007,Tong2016}.
	\item $a=4/3$ in the case without disk wind while $a>4/3$ on the contrary. In general, $a=5/3$ for the strongest wind loss so that $4/3<a<5/3$ if the loss is weak \citep{Pringle1991,Cannizzo2009,Shen2012,Liu2015,Lin2021}.
\end{enumerate}
In this work, we mainly discuss the influence of disk wind.

With the mass flow rate decreasing and the size expanding, the disk usually starts at a slim or thick disk phase, then translates to a thin disk and ends at a advection-dominated accretion flow \citep{Liu2015,Xu2019}.
With the intention of studying the interaction between the fallback disk and the NS, we define three radii.
First, the inner radius of the disk is $R_{\rm in} = \xi R_{\rm A}$, where $\xi=1$ in this paper, and $R_{\rm A}$ is the traditional Alfv\'en radius for spherical accretion 
\begin{equation}
    R_{\rm A}=\left(\frac{\mu^4}{2GM_{\rm NS}\dot{M}_{\rm in}^2}\right)^{1/7},
\end{equation}
where $G$ is the gravitational constant and $\dot{M}_{\rm in}$ is the mass flow rate at the inner edge of the disk.
The second is the corotation radius $R_{\rm c}$ in which the Keplerian angular velocity $\Omega_{\rm K}(R)$ of the disk equals the angular velocity $\Omega_{\rm s}$ of the NS
\begin{equation}
    R_{\rm c}=\left(\frac{GM_{\rm NS}}{\Omega_{\rm s}}\right)^{1/3}.
\end{equation}
Lastly, the light cylinder radius is defined as where the corotation velocity is equivalent to the speed of light $c$
\begin{equation}
    R_{\rm lc}=\frac{c}{\Omega_{\rm s}}.
\end{equation}

When $R_{\rm in}<R_{\rm c}$, accretion occurs and the NS gains angular momentum at a rate
\begin{equation}
    N_{\rm spin-up}=\dot{M}_{\rm acc} (GM_{\rm NS} R_{\rm in})^{1/2},
\end{equation}
to spin up, where $\dot{M}_{\rm acc}$ is the accretion rate onto the NS.
While $R_{\rm c}<R_{\rm in}<R_{\rm lc}$, the propeller phase is triggered and the material in the inner edge of the disk is ejected, so the NS spins down with a torque
\begin{equation}
    N_{\rm spin-down}=-\dot{M}_{\rm in} (GM_{\rm NS} R_{\rm in})^{1/2}.
\end{equation}
When $R_{\rm in}>R_{\rm lc}$, there is no interaction between the disk and the NS, only magnetic dipole radiation operates.

\section{Results}

\subsection{Model without disk wind}\label{no_wind}

\begin{figure}
	\plotone{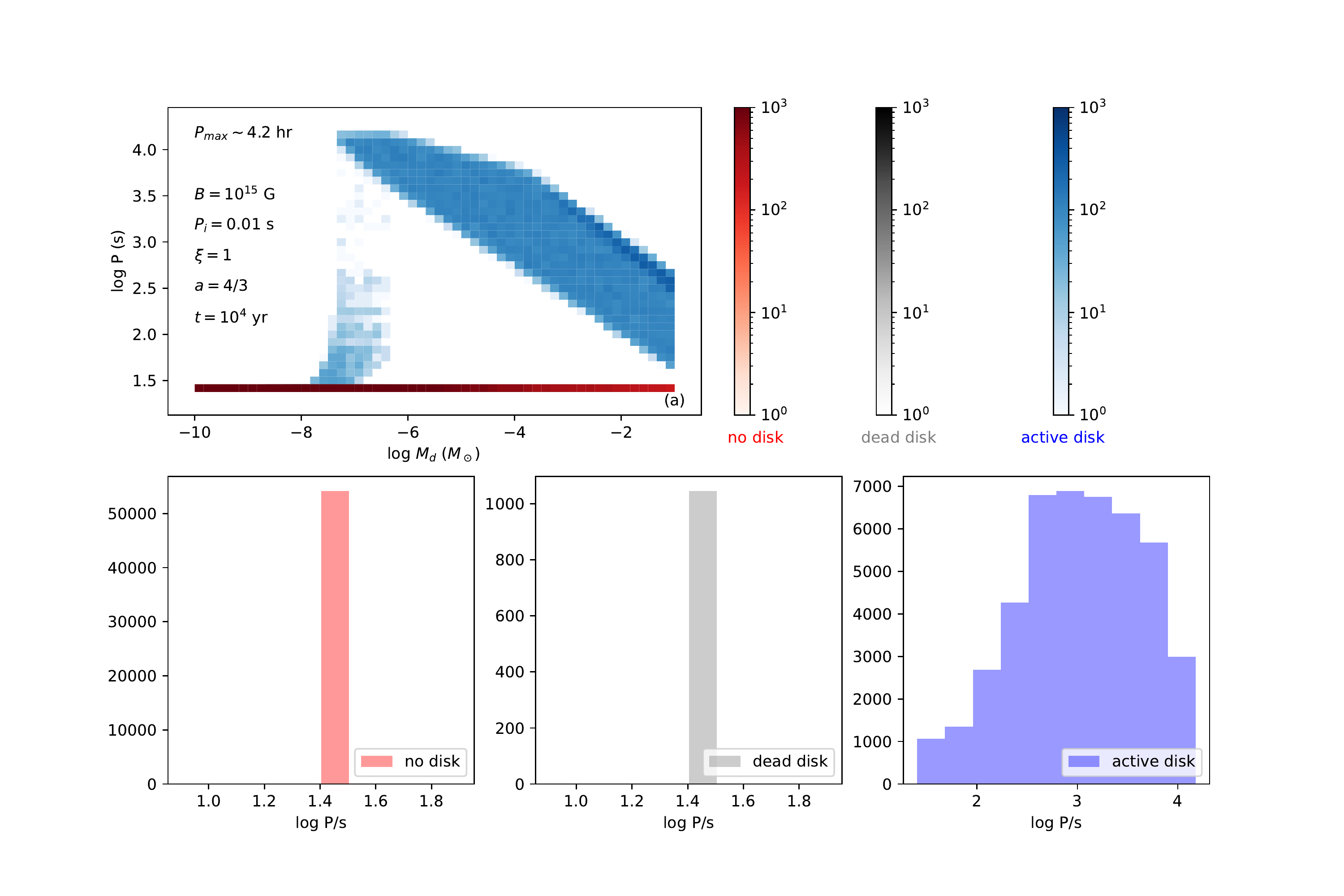}
	\caption{Top panel: distribution of spin period at age of $10^4$ yr against $M_{\rm d}$ with the magnetic field $B$ taken to be $10^{15}$ G. The blue and gray colors demonstrate the cases that the disks are in active and dead state, respectively, while the red colors mean that the disk cannot form. The color depth describes the number of NSs in each bin, which is indicated in the color bar. And the color bars are shared in all the figures of MC results in this paper if there is no color bar exhibited beside the panels.
	Lower panels: the histograms of no disk cases, dead disk cases and active disk cases from left to right.
		\label{fig:b15_yr4_a4}}
\end{figure}

\begin{figure}
	\plotone{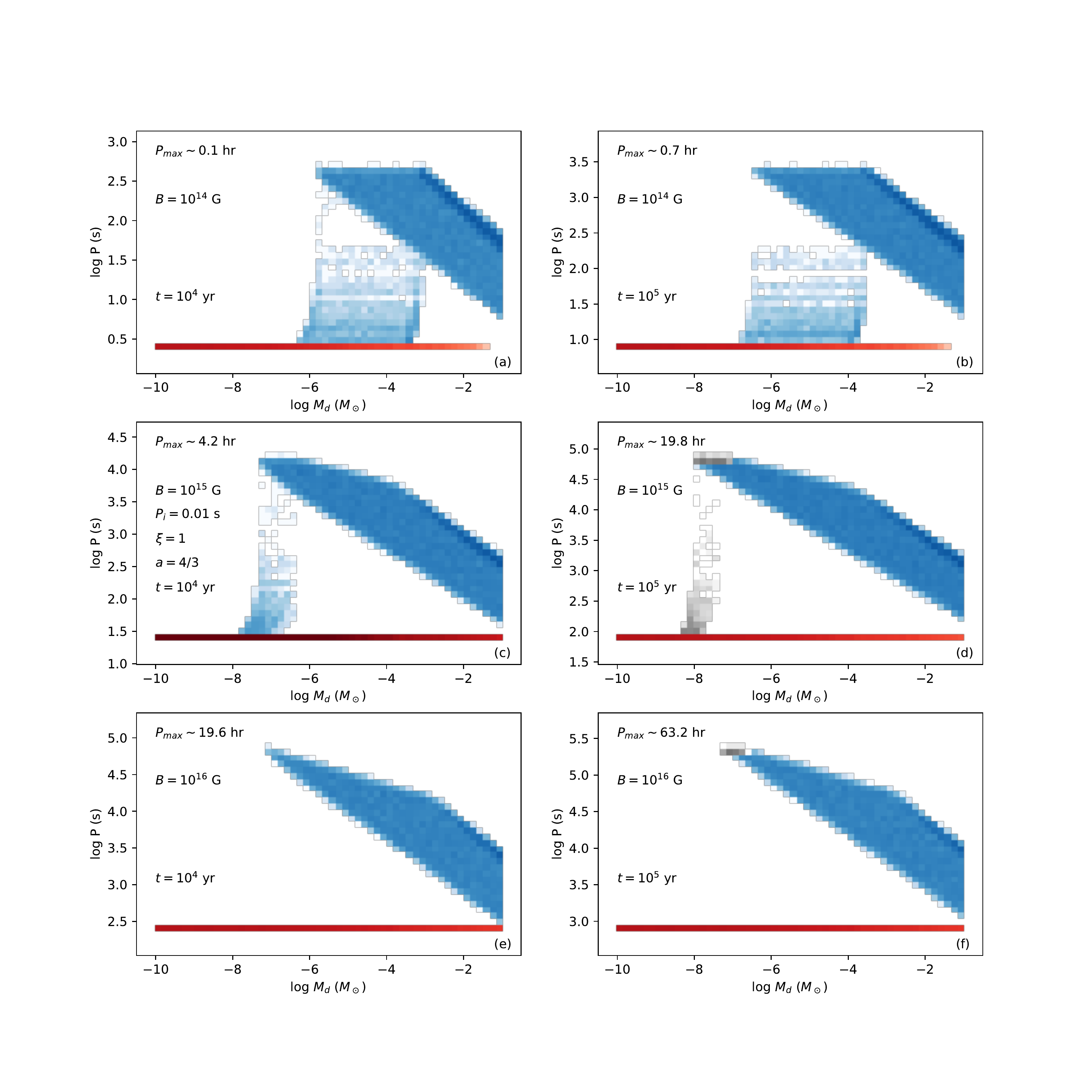}
	\caption{Distribution of spin period against $M_{\rm d}$ in 6 groups of MC simulations with $a=4/3$.
		\label{fig:b_yr_a4}}
\end{figure}

We first set up $a=4/3$ in our Monte-Carlo simulation without considering disk wind.
In each case, the torque supported by the fallback disk comes into operation at $t_{\rm f}$ and ends at the age of a magnetar (taken to be $10^4$ or $10^5$ yr in this paper) or when the disk enters "dead disk" state where there is no interaction between the NS and the disk \citep{Xu2019}.
The initial mass $M_{\rm d}$ and the outer radius $R_{\rm f}$ of the disk are randomly distributed in the range of $[10^{-10},10^{-1}] \ M_{\odot}$ and $[R_{\rm NS},10^6R_{\rm S}]$, where $R_{\rm NS}=10^6$ cm and $R_{\rm S}\approx 4 \times 10^5$ cm are the radius and the Schwartzschild radius of a 1.4 $M_{\odot}$ NS.
And in each sample we simulate the evolution of $10^6$ NSs.
Top panel of Figure \ref{fig:b15_yr4_a4} shows the distribution of the spin periods $P$ at $10^4$ yr versus the initial disk mass $10^{-10} M_\odot \leq M_{\rm d} \leq 10^{-1} M_\odot$\footnote{Another free parameter is the initial outer radius of the fallback disk $R_{\rm f}$, but the the distribution of the spin periods are not very sensitive to it as showed in the left panel of Figure 2 in XL19, so we won't exhibit the results versus $R_{\rm f}$ in this paper.}.
Other parameters are listed in the figure, which are the magnetic field of the NS $B=10^{15}$ G, 
the initial spin period of the NS $P_{\rm i} = 0.01$ s and the correction factor of the inner disk radius $\xi=1$. The maximum spin period $P_{\rm max}$ is about 4.2 hr.
The blue and gray colors demonstrate the cases that the disks are in active and dead state, respectively, while the red colors mean that the disk cannot form.
The color depth describes the number of NSs in each bin, which is indicated in the color bar. And the color bars are shared in all the figures of MC results in this paper if there is no color bar exhibited beside the panels.
The lower panels show the histograms of no disk cases, dead disk cases and active disk cases from left to right, respectively.
Then we show the results of 6 groups MC simulations with $a=4/3$ in Figure \ref{fig:b_yr_a4}. From up to bottom panels, the magnetic fields are $B=10^{14}$, $10^{15}$ and $10^{16}$ G, respectively. 
It shows that $P_{\rm max}$ becomes larger for stronger $B$, which is consistent the the results in XL19.
This implies that the NS has reached the equlibrium spin period 
\begin{equation}
	P_{\rm eq}=2^{11/14} \pi (GM)^{-5/7} \xi^{3/2} \mu^{6/7} \dot{M}_{\rm in}^{-3/7}. \label{peq}
\end{equation}
The age of the systems are $10^4$ yr in the left panels and $10^5$ yr in the right panels.
NSs spin down to much longer spin periods at $10^5$ yr but the disks evolve to dead state in more systems.

\begin{figure}
	\plotone{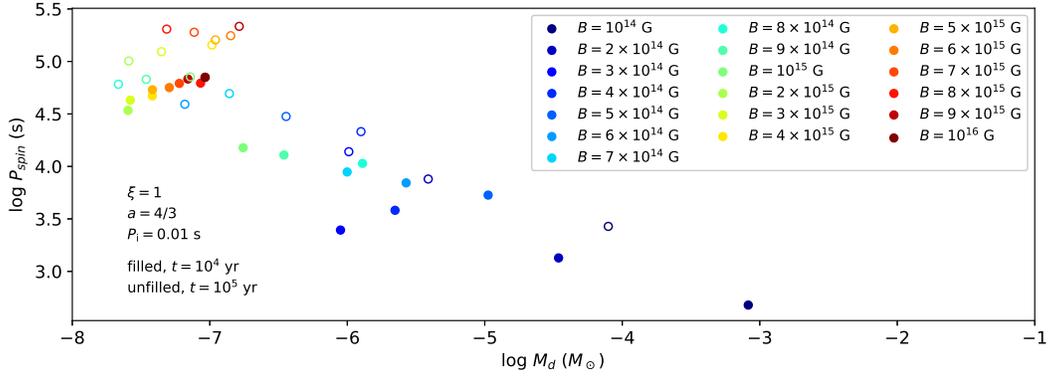}
	\caption{The maximum spin periods against $M_{\rm d}$ in the model without disk wind ($a=4/3$). The filled and unfilled circles  correspond to the case with $t=10^4$ and $10^5$ yr. The colors indicate the magnetic fields of the NS.
		\label{fig:pmax_a4}}
\end{figure}

We do more simulations with $10^{14}~{\rm G}\leq B \leq 10^{16}~{\rm G}$ and pick out $P_{\rm max}$ in each MC simulation group drawing in Figure \ref{fig:pmax_a4}.
The filled and unfilled circles correspond to the case with $t=10^4$ yr and $10^5$ yr,  respectively. The colors indicate the magnetic field of the NS, which are demonstrated in the legend.
It shows that $P_{\rm max} \approx 19.6$ hr at $10^4$ yr and $P_{\rm max} \approx 63.2$ hr at $10^5$ yr in the model without disk wind when $B=10^{16}$ G.

\subsection{Model with disk wind}\label{wind}

\begin{figure}
	\plotone{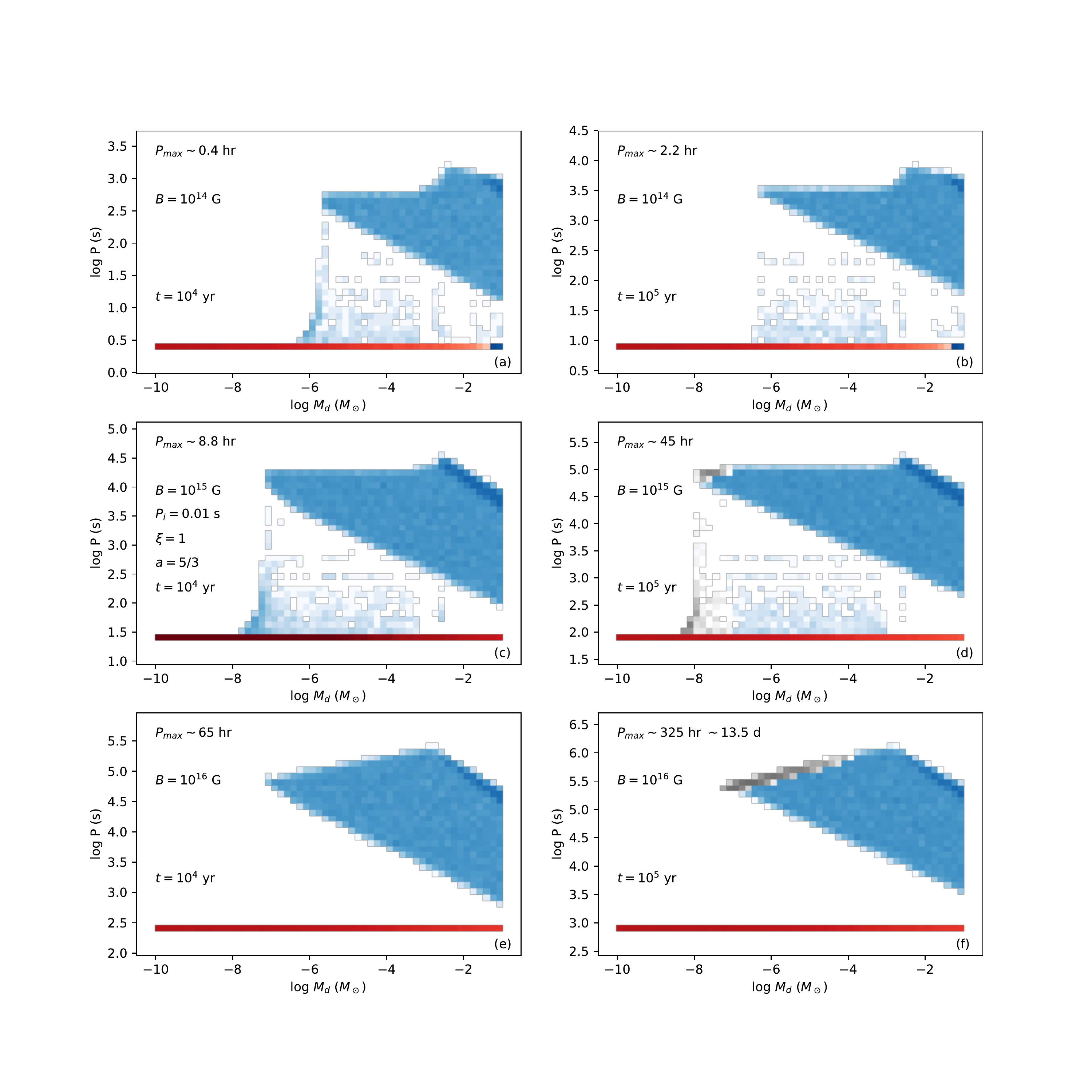}
	\caption{Same as Figure \ref{fig:b_yr_a4} with $a=5/3$.
		\label{fig:b_yr_a5}}
\end{figure}

\begin{figure}
	\plotone{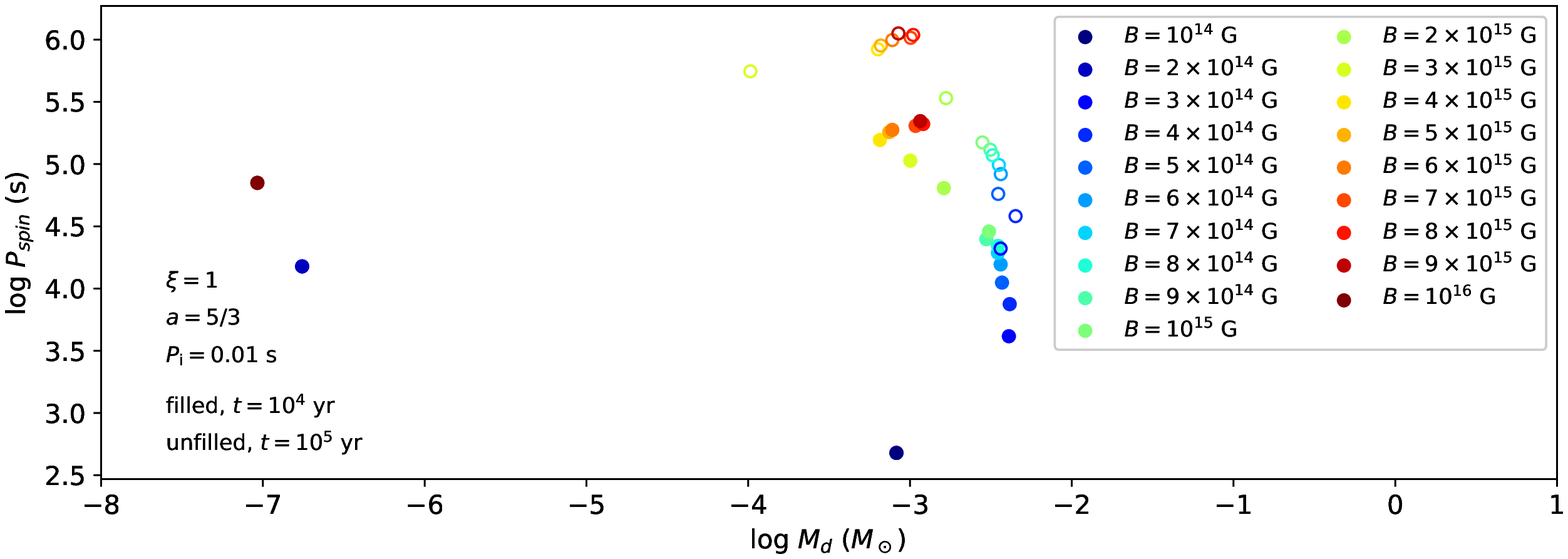}
	\caption{Same as Figure \ref{fig:pmax_a4} but with $a=5/3$.
		\label{fig:pmax_a5}}
\end{figure}

We vary the value of $a$ to be $5/3$ in this section to see how the disk wind can influence the spin evolution of the NS. 
The results are showed in Figure \ref{fig:b_yr_a5} and \ref{fig:pmax_a5}. 
They show that the NS can spin down to a much long spin period under the action of disk wind. 
And the maximum values are also got when $B=10^{16}$ G, which is $P_{\rm max} \approx 65$ hr at $10^4$ yr and 
$P_{\rm max} \approx 325$ hr ($\approx 13.5$ day) at $10^5$ yr.

\section{Discussion}

\subsection{Range of the power law index $a$}

\begin{figure}
	\plotone{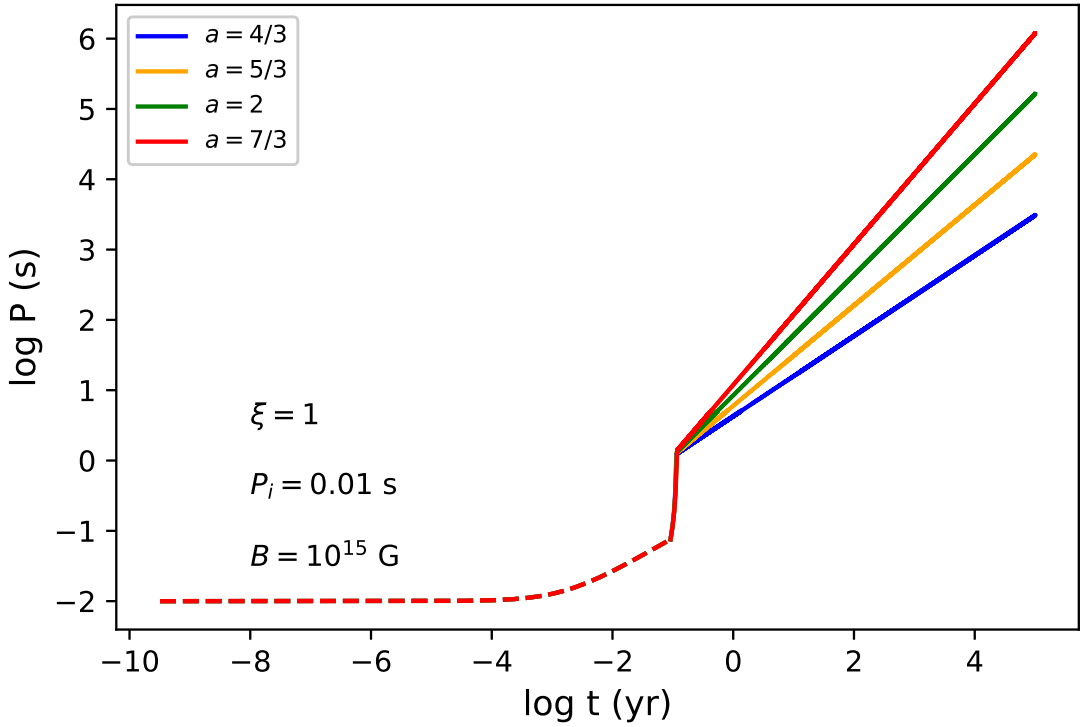}
	\caption{Spin evolution with the power law index $a$ taken to be $4/3$, $5/3$, $2$ and $7/3$. In all cases $M_{\rm d}=10^{-3} M_\odot$ and $R_{\rm f}=10^4 R_{\rm S}$. 
	The four curves coincide before $t \sim 10$ yr, where the disk haven't form (the dashed part) or is in propeller phase (the solid part). When entering equlibrium state, the curves seperate and $P_{\rm eq}$ becomes larger as $a$ increases.
	\label{fig:b15_t5_a}}
\end{figure}

There are some debates on the value of $a$ in disk wind model.
\cite{Beniamini2020} provides that $a=\frac{28(p+1)}{3(7+2p)}$ and $0\leq p \leq 1$ \citep{Blandford1999} for the classical thin disk model, and thinks it possibly associated with radiatively-inefficient accretion
flows (RIAFs).
In fact the rotation of the disk is thought to be sub-Keplerian when the inner region of the disk becomes RIAFs (slim/thick disk) \citep{DallOsso2016,Xu2017,Gao2021}, 
so the relationship between the inner disk radius and the mass transfer rate should be $R_{\rm in} \propto \dot{M}_{\rm in}^{-1/7}$ \citep{DallOsso2016} or even more complex form \citep{Xu2017}, rather than $R_{\rm in} \propto \dot{M}_{\rm in}^{-2/7}$. And based on the deducing process in \cite{Beniamini2020} one can get $a=\frac{28(p+1)}{3(7+p)}$ for the slim or thick disk model. 
Where $p=0$ means there is no wind \citep{Blandford1999} and $0<p \leq 1$ corresponding to the disk wind case, i.e., $4/3 \leq a \leq 7/3$. 
Combining Equations \eqref{mdot} and \eqref{peq} we have $P_{\rm eq} \propto \dot{M}^{-3/7} \propto t^{3a/7}$. 
We show some cases of spin evolution path with $a=4/3$, $5/3$, $2$ and $7/3$ in Figure \ref{fig:b15_t5_a}. In each curve the dashed and solid parts indicate the evolutionary stages before and after the fallback disk formed, respectively.
The four curves coincide before $t \sim 10$ yr, where the disk haven't form or is in propeller phase. 
When entering equilibrium state, the curves separate and $P_{\rm eq}$ becomes larger as $a$ increases.

``Disk winds'' are thought to be observed in four NS-systems, which are
Her X-1 \citep{Kosec2020,Nixon2020},
GX 13+1 \citep{Ueda2004}, IGR J17480-2446 \citep{Miller2011} and IGR J17591-2342 \citep{Nowak2019}.
In the first three systems, outflows are probably from the inner disk in super-Eddington accretion phase, which may be explained by the mass loss at the spherization radius because the Eddington-limited accretion is enabled \citep{Shakura1973,Lipunova1999,Grebenev2017}. Only in the fourth system, emission radii ranges from 1000 to 200,000 km, which is consistent with the disk wind scenario proposed in \cite{Emmering1992}. 
It's still not clear about the driving mechanism and energy budget of disk wind.
The hard X-ray radiation and magnetic fields on the disk are the favored explanations \citep{Drew2000,Blandford1982} in nonmagnetic accretor systems, i.e., black holes, nonmagnetic cataclysmic variables and young stellar objects.
But in the NS-systems, the disorganized magnetic field lines on the disk will be arranged and reconnect with the lines of the NS.
So the material is possibly constrained by the reconnected lines and hard to escape from the disk.

Since $0\leq p \leq 1$ is adapted to the disk around a black hole, which may be different in an NS system.
And in observation, none of disk wind phenomenon was detected in isolated NS-systems\footnote{The four NS-systems thought to be observed ``Disk winds'' are all binaries.}, which may be because the fallback disk mass is too small to generate strong and sustained wind.
Then from the geometrical feature, a slim disk is more likely to generate disk wind than a thin disk, but its lifetime\footnote{One can get that the lifetime of slim disk is $\lesssim 1$ yr from the Appendix B in XL19} is very short, i.e., even if the disk wind were strong in the slim disk phase, it can be ignored.
On the other hand, one can know from Figure \ref{fig:b15_t5_a} that an NS can easily spin down to tens of thousand seconds within a few thousands years under the action of strong disk wind, while only one very slow isolated NS (1E 1613) was found in observation.
Therefore, disk wind in isolated NS-systems might be very weak, which can be even ignored, and the maximum value of $a$ is possibly not as large as $7/3$ while $5/3$ is more physical. 
So the maximum spin periods of isolated NSs are $\sim 63$ hr without considering disk wind and $\sim 13.5$ day with disk wind at $10^5$ years.

\subsection{Periodicity in FRB 180916 and 121102}

\begin{figure}
	\plotone{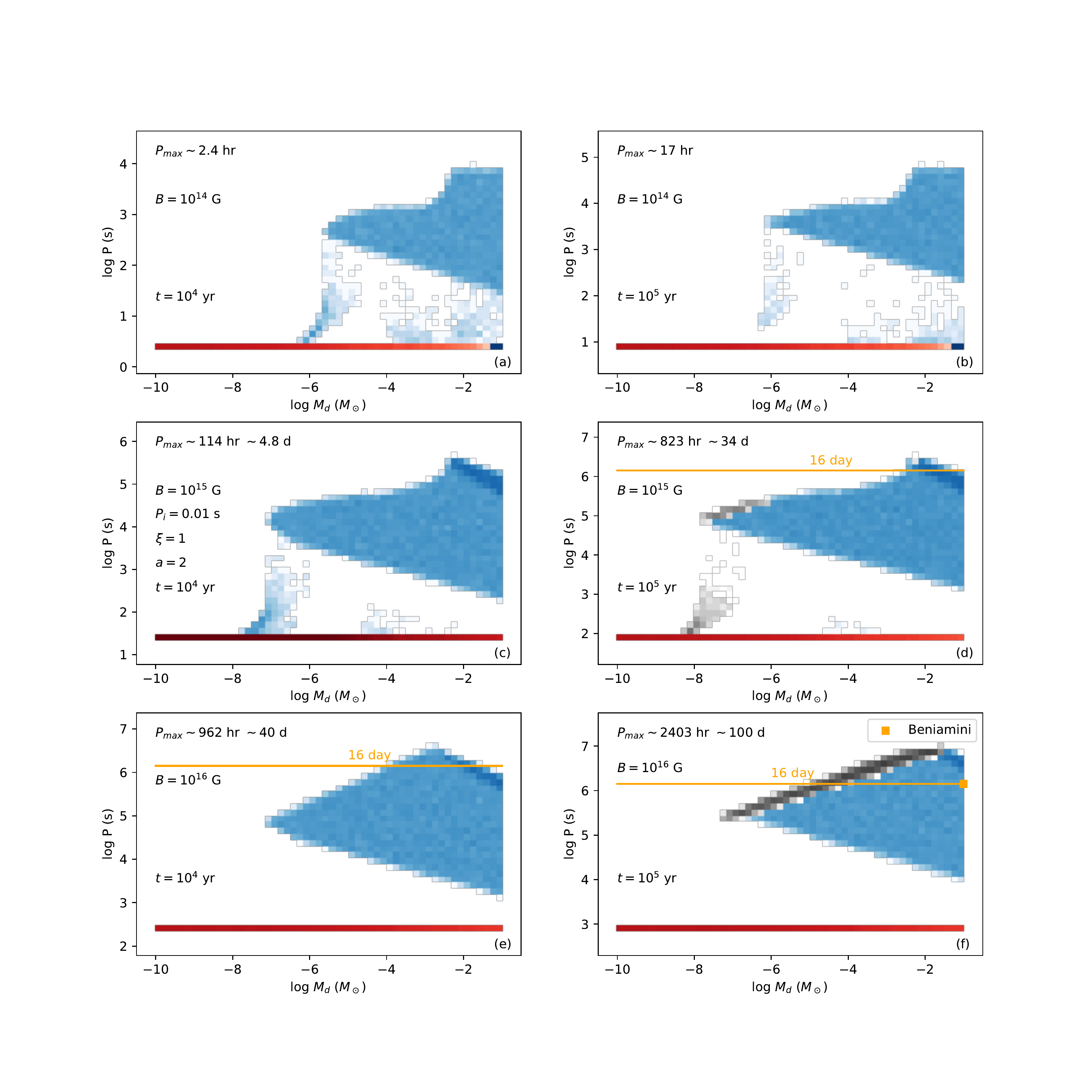}
	\caption{Same as Figure \ref{fig:b_yr_a4} with $a=2$. The orange solid line respresents the spin period of $16$ day. The orange square corresponds to the results in \cite{Beniamini2020}.
		\label{fig:b_yr_a6}}
\end{figure}

\begin{figure}
	\plotone{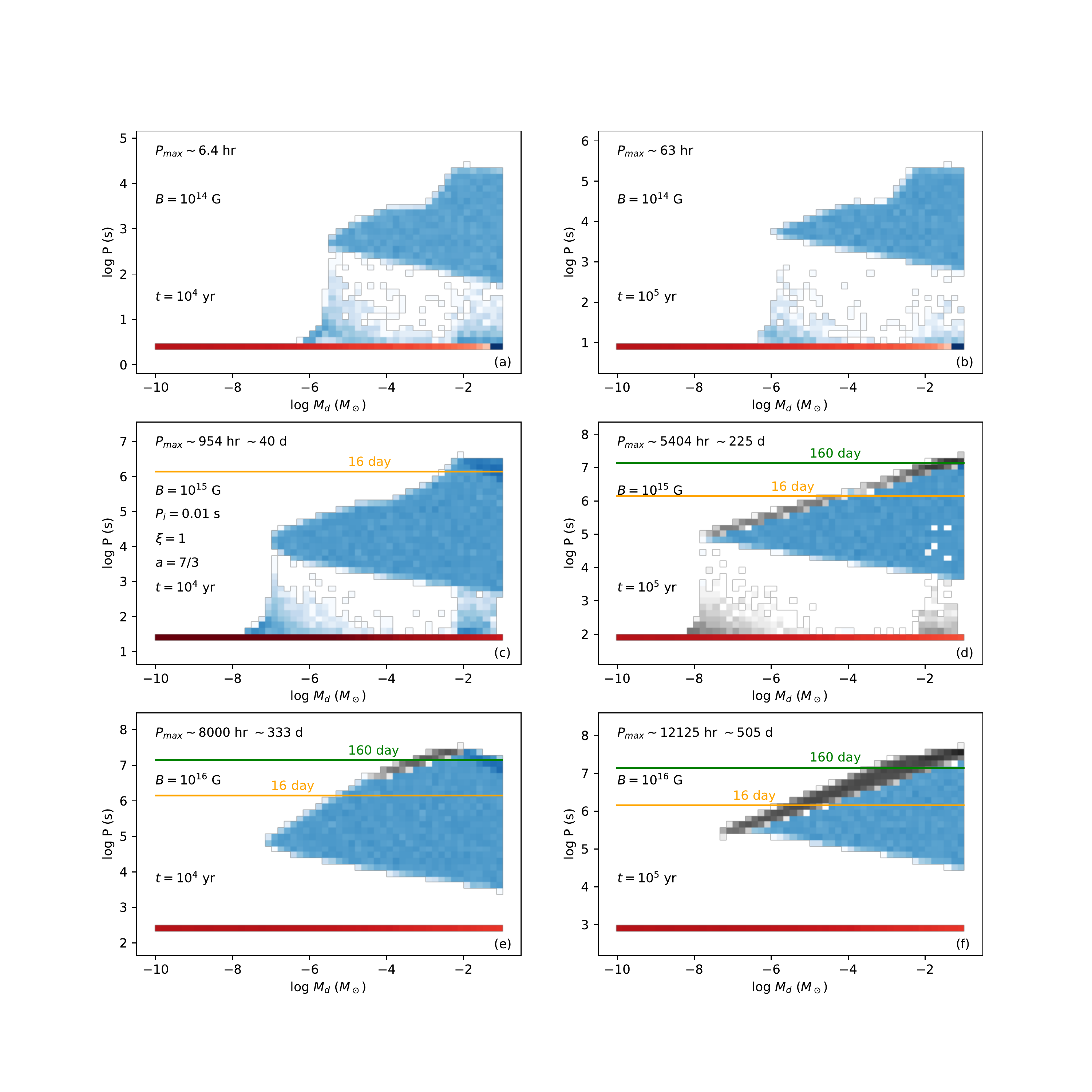}
	\caption{Same as Figure \ref{fig:b_yr_a4} with $a=7/3$. The orange and green solid lines respresent the spin period of $16$ day and $160$ day, respectively.
		\label{fig:b_yr_a7}}
\end{figure}

\begin{figure}
	\plotone{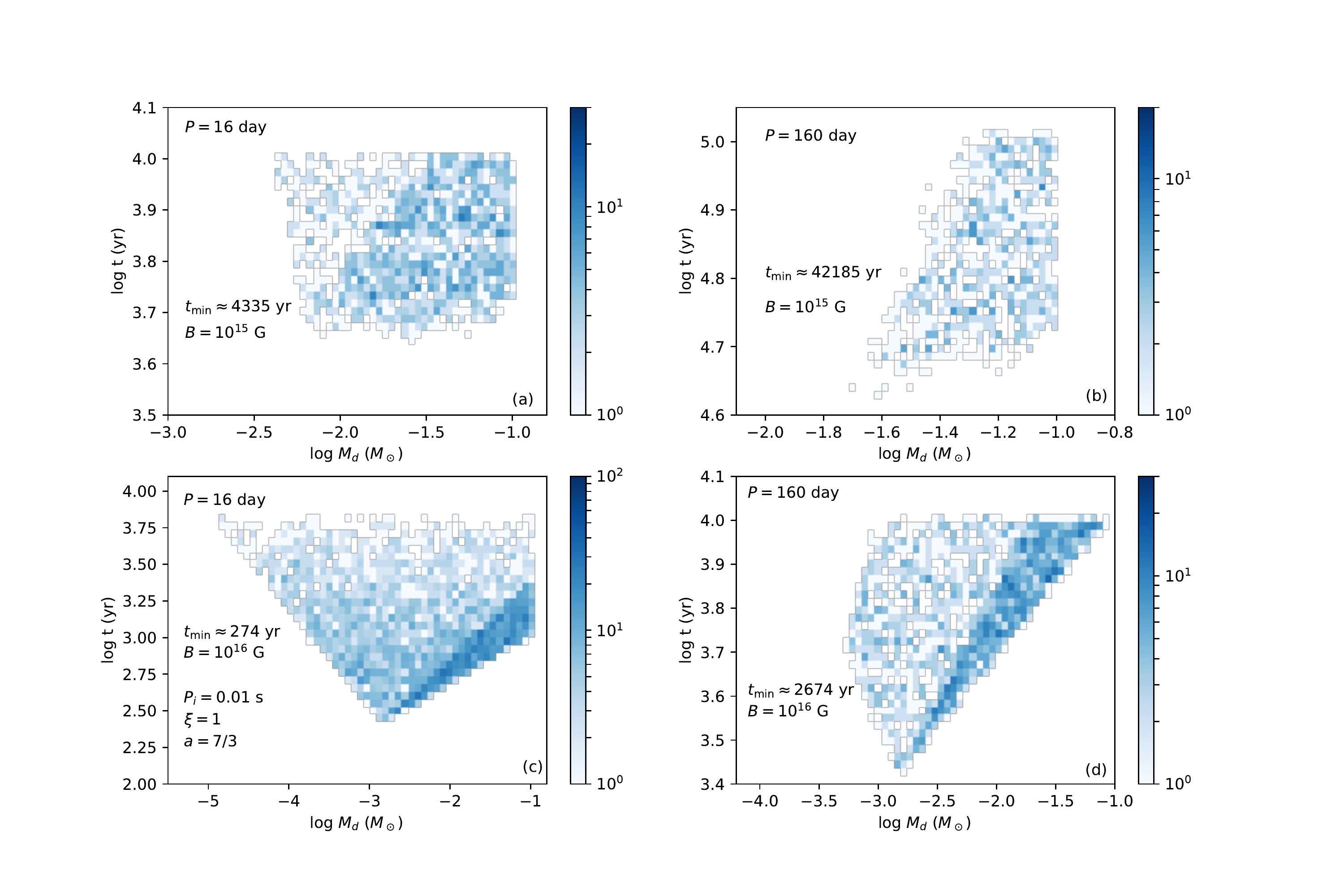}
	\caption{Distribution of NS age against $M_{\rm d}$ when the spin periods reach 16 day (the left panels) or 160 day (the right panels) in the cases with $a=7/3$.
		\label{fig:p16_p160_age}}
\end{figure}

As showed in above, the maximum spin period of an isolated NS/magnetar is $\sim 13.5$ day with a power law index $a \leq 5/3$, which is
widly acceptable and in accordance with theory and observation. 
Since the possibility of the ultra-strong disk wind cases with $a>5/3$ can't be ruled out,
we also do the MC simulations with $a=2$ and $a=7/3$. 
We show all the results of 24 groups of MC simulation in Figure \ref{fig:b_yr_a4}, \ref{fig:b_yr_a5}, \ref{fig:b_yr_a6} and \ref{fig:b_yr_a7} with $a=4/3$, $5/3$, $2$ and $7/3$ respectively. 
The orange and green solid lines represent the spin period of $16$ day and $160$ day, respectively.
It shows that the spin period can be larger than $16$ day or even $160$ day in some groups when $a>5/3$. 
The orange square in lower right panel of Figure \ref{fig:b_yr_a6} corresponds to the results in \cite{Beniamini2020}, which is covered by our results.
The parameter spaces with $P>16$ day and $P>160$ day are exhibited in Table \ref{table:p_16_160}.
Figure \ref{fig:p16_p160_age} displays the distribution of NS age against $M_{\rm d}$ when the spin periods reach 16 day (the left panels) and 160 day (the right panels) in the case with $a=7/3$. 
It shows that an NS can spin down to 16 day within a few hundred years, 
which is much shorter than the age of a magnetar, 
if the magnetic field is as extremely strong as $10^{16}$ G and the power law index is as awfully large as $7/3$.
So if one can give a physical meaning of $a>5/3$, the long spin period of magnetars may connect to the repeating FRBs and perhaps some other interesting results of magnetars can be got.

\begin{deluxetable}{ccccc}
\tablewidth{0pt}
\tablecaption{Parameter spaces with $P>16$ day and $P>160$ day \label{table:p_16_160}} 
\tablehead{\colhead{$a$} & \colhead{$B$/G} & \colhead{t/yr} & \colhead{$M_{\rm d}/M_\odot$ ($P>16$ day)}  & \colhead{$M_{\rm d}/M_\odot$ ($P>160$ day)}
}
\startdata
\multirow{3}*{$2$} & $10^{15}$ & $10^5$ & $[3 \times 10^{-3}, ~ 5 \times 10^{-2}] $ & - \\
\cline{2-5}
& \multirow{2}*{$10^{16}$} & $10^4$ & $[2 \times 10^{-4}, ~ 2 \times 10^{-2}]$ & - \\
\cline{3-5}
 &  & $10^5$ & $[10^{-5}, ~ 10^{-1}]$ & - \\
\hline
\multirow{4}*{$7/3$} & \multirow{2}*{$10^{15}$} & $10^4$ & $[4 \times 10^{-3}, ~ 10^{-1}]$ & - \\
\cline{3-5}
 &  & $10^5$ & $[6 \times 10^{-5}, ~ 10^{-1}]$ & $[2 \times 10^{-2}, ~ 10^{-1}]$ \\
\cline{2-5}
 & \multirow{2}*{$10^{16}$} & $10^4$ & $[7 \times 10^{-6}, ~ 10^{-1}]$ & $[6 \times 10^{-4}, ~ 7 \times 10^{-2}]$ \\
\cline{3-5}
 &  & $10^5$ & $[2 \times 10^{-6}, ~ 10^{-1}]$ & $[6 \times 10^{-4}, ~ 10^{-1}]$ \\
\hline
\enddata
\end{deluxetable}

If 16 days or more is indeed a period of a magnetar, 
it is an open question what kind of mechanism and process can produce repeating FRBs. 
The slowest isolated pulsar, 1E 1613, in the supernova remnant RCW 103 has a spin period $\sim 6.67\,\unit{hr}$ \citep{De2006} and can be explained by a magnetar interacting with a fallback disk \citep{Xu2019}. When the fallback disk is in propeller state, the period of the magnetar will become long. Although the matter on the disk is thrown out by the centrifugal force, there is still a small part of the material which may fall into the poles of the NS along the magnetic field line. 
However, if the periodicity of a FRB is caused by the rotation of the magnetar itself, it may be more possible for short period magnetars with a more effective accretion. This process of producing FRBs is similar to an impact between an asteroid and a magnetar because of accretion \citep{gen20}.

\section{Summary}

The period of some repeating FRBs, FRB 180916 and FRB 121102 may be from a binary system, the precession or the spin of an NS\footnote{The binary scenario seems disfavored by \citep{pas20} and the simplest  precession models seem inconsistent with polarization measurements \citep{Nimmo2021}}.
In this work, we explore the possible longest spin period of an isolated NS interacting with a fallback disk with and without disk wind to explain the periodic activity of FRB 180916. 
Our results show that $P_{\rm max} \sim 63$ hr without considering disk wind ($a=4/3$) and $P_{\rm max}  \sim 13.5$ day with a classical parameter of disk wind ($4/3<a \leq 5/4$) if the magnetic field of the NS is as extremely strong as $10^{16}$ G and does not decay within $10^5$ years.
However we can obtain the spin period of hundreds of days with a large index of mass flow rate.
Yet, the following arguments suggest such strong disk wind ($a > 5/3$) around a highly magnetic NS may be unphysical:
\begin{enumerate}
    \item From the geometrical feature, a slim disk (or RIAF from \cite{Beniamini2020}) is more likely to generate disk wind than a thin disk, but its lifetime is very short, i.e., even if the disk wind were strong in the slim disk phase, it can be ignored,
    \item The fallback disk mass is usually too small ($< 0.1 M_{\odot}$) to generate strong and sustained wind,
    \item None of disk wind phenomenon was detected in isolated NS-systems,
    \item The slowest isolated pulsar and accreting X-ray binary pulsar are 1E 161348+ 5055 with $P_{\rm spin} \approx 24000 s$ \citep{De2006} and AX J1910.7+0917 with $P_{\rm spin} \approx 36200 s$ \citep{Sidoli2017}, respectively. But they are all detected in X-ray and no radio observation reported, which maybe indicate that they are in the "grave yard" of radio pulsars.
\end{enumerate}
We find that the possible maximum spin period of an isolated NS is $\sim 13.5$ day\footnote{Since the slowest pulsar in isolated and binary systems known by now are both detected in X-ray, as well as considering the "grave yard" of radio pulsars, we suggest this pulsation should be in X-ray rather than radio.} with $a \leq 5/3$.
Therefore, the periodic activity of these repeating FRBs may be not due to the NS spin if standard fallback disk models are invoked.

\acknowledgements 
We thank the anonymous referee for helpful comments and suggestions.
We thank Victoria Kaspi and Xuan Fang for helpful discussions.
This work was supported by the Natural Science Foundation of China under grant No. 11773015, 11833003, 11933004, 11988101 and 12063001. 
Y.P.Y is supported by National Natural Science Foundation of China grant No. 12003028 and Yunnan University grant No.C176220100087.

\newpage

\clearpage

\label{lastpage}
\end{document}